# Alternative Models of Zebra Patterns in the Event on June 21, 2011

L.V. Yasnov[1,2], G.P. Chernov[3]

[1] St.-Petersburg State University, St.-Petersburg, 198504, Russia (e-mail: leonid.yasnov@mail.ru)

[2] St.-Petersburg branch of Special Astrophysical Observatory, St.-Petersburg, 196140, Russia

[3] Pushkov Institute of Terrestrial Magnetism, Ionosphere and Radio wave propagation of Russian Academy of Sciences (IZMIRAN), Moscow, Troitsk, 108840, Russia

**Abstract** The analysis of the spectral characteristics of the burst radio emission on June 21, 2011 was carried out on the basis of an improved methodology for determining harmonic numbers for the corresponding stripes of the zebra structure. By using the parameters of the zebra structure in the time-frequency spectrum and basing on the double plasma resonance model, the magnetic field and its dynamics, electron density, and the time variation of the distance between the stripes with harmonics $s = 55$ and $56$ and adjacent stripes near the frequency $\approx 183$ MHz have been determined in the burst generation region. The relationships between the scale characteristics of the field and the density along and across the axis of the power tube and their dependence on time have been also determined. The field obtained (1.5 G for the first harmonic and 0.75 G for the second harmonic of the plasma frequency) turned out to be so small that, firstly, it fails to explain the dynamic features of the spectrum based on MHD waves, and secondly, it results in large values for plasma betta ($>1$). Other possible difficulties of the generation mechanism of bursts with zebra pattern based on the double plasma resonance are also noted. Another possible mechanism - with whistlers explains qualitatively the main observational characteristics of this zebra. The magnetic field required in this case is about 4.5 G, and the plasma betta is 0.14, which fully corresponds to the coronal conditions.

# 1. Introduction

The zebra pattern (ZP) in the form of regular stripes in emission and absorption in the dynamic spectrum of solar radio bursts has been studied during five cycles of solar activity. Their basic observational properties are represented in many reviews and monographs (Slottje, 1981; Kuijpers, 1975; Chernov, 2006, Chernov, 2011). Theoretical models of ZP generation have been simultaneously studied, and at the present time more than ten mechanisms have been proposed. The mechanism based on the double plasma resonance (DPR) is the one most widely discussed in literature. The mechanism assumes that the emission occurs at the upper hybrid frequency ($\omega_{UH}$), when it becomes equal to the integer number (*s*) of the electronic cyclotron harmonics ($\omega_{Be}$):

$$\omega_{UH} = (\omega^2_{Pe} + \omega^2_{Be})^{1/2} = s\omega_{Be} , \qquad (1)$$

where $\omega_{Pe}$- the electron plasma frequency, with the condition that $\omega_{Be} \ll \omega_{Pe}$ (Kuijpers, 1975; Zheleznykov and Zlotnik, 1975; Kuijpers (1980); Mollwo, 1983; 1988; Winglee and Dulk, 1986) Zheleznykov et al. 2016. In these papers it is usually postulated that the mechanism always works, if there are fast particles in the magnetic trap. However, the mechanism encounters some difficulties in explaining the dynamics of Zebra-stripes (a sharp change in the frequency drift of stripes, the large number of harmonics, the frequency splitting of stripes, their superfine structure in the form of millisecond spikes). Therefore the works aimed at improving the model (Karlický *et al*., 2001; LaBelle *et al*., 2003; Kuznetsov and Tsap, 2007*)*, as well as creating new ones began to appear.

An alternative mechanism of the interaction of plasma waves (*l*) with the whistlers (*w*) was almost simultaneously proposed: $l + w \rightarrow t$ (Chernov, 1976; 1996). In this model the above-mentioned thin effects of zebra stripes are explained by the quasi-linear effects of interaction of fast particles with whistlers. The mechanism with whistlers became its natural development after its being applied by Kuijpers (1975) for the fibers (fiber bursts), when in some events there was observed a continuous transition of zebra stripes into the fibers and backward. The most important details of this mechanism are represented in the section Discussion.

In the fundamental paper on the DPR mechanism by Zheleznykov and Zlotnik (1975) the relative width of the increment zone in the hybrid band was shown to be extremely narrow: $\delta\omega/\omega_{Be} \approx 2.5\times10^{-4}$. Such a value is to be obtained provided the dispersion of the particle beam velocities during the estimations is not taken into account as an infinitesimal quantity. In the paper by

Alternative Models. . .

Benáček, Karlický and Yasnov (2017) the effectiveness of the DPR mechanism was shown to be limited when taking into account the beam velocities dispersion for hot particles as well as the temperature of cold plasma (for more details see Discussion).

In several studies, the formation of ZP is associated with the diffraction of radio waves by coronal inhomogeneities (Laptuhov and Chernov, 2006; 2012; Barta and Karlicky, 2006) or the interference between the direct rays and the reflected ones from the inhomogeneities (Ledenev, Yan, and Fu, 2006). A completely new approach based on explosive instability was proposed in the work by Fomichev, Fainshtein, Chernov (2009). This mechanism may prove to be very promising, since the frequency separation between the zebra stripes in this mechanism is correlated with the frequency of ionsound waves and does not depend on the relationship between plasma and gyrofrequency. The significance of these models (and a number of others) is still worth clarifying. Therefore in this work we have compared the possibilities of these two old alternative models by taking the ZP example in the event of 21 June 2011.

## 2. The DPR mechanism

It was shown in (Yasnov, Benáček and Karlický, 2017) that, when the line of sight is along the axis of a source, with the density varying with height and radius, the dimensions of the emitting region strongly depend on the density gradient along the axis of the source (along the height in the case of a vertical location of the source relative to the photosphere). So for the height scale along the axis of the source with respect to the half-width of the source in the range 1÷3, the size of the source, due to the directivity of the emission, cannot exceed 0.006-0.018 fractions of the half-width of the source (see Figure 2 in this paper). This leads to the fact that the brightness temperature must take on very high values so that the flux would correspond to the observed one. It is unlikely that such temperatures (up to $10^{17}$ K) could be possible. But the loops can consist of either several unresolved filaments or a single filament that does not fill the whole volume of the loop (Aschwanden et al. 2008). This makes the registration of emission even more improbable when observing a radiating region along its longitudinal axis.

In this regard, a model of burst generation sources with a zebra structure in the form of a plasma tube was proposed in (Karlický and Yasnov, 2018a). In this model, the density is specified as:

$$N = N_0 \, \mathrm{e}^{-r^2 - \frac{h}{L_{\mathrm{nh}}}}, \qquad (2)$$

where $L_{nh}$ is the density scale along the height. The density scale along $r$ (close to the radius of the tube), without loss of commonality, can be taken equal to 1. The magnetic field model is set in the same way:

$$B = B_0 e^{-r^2/L_{br}^2 - h/L_{bh}}, \quad (3)$$

where $L_{br}$ and $L_{bh}$ — are correspondingly the field scales along the radius $r$ and along the height $h$ of the tube. In such a model, as shown in (Karlický and Yasnov, 2019), in order to generate isolated zebra stripes, it is necessary to fulfill the following relation for the dimensionless field and density scales:

$$L_{bh} = L_{br}^2 L_{nh}. \quad (4)$$

Emission regions are rather narrow formations – the magnetic tubes. The analysis of possible conditions in such tubes shows that the DPR layers have a very elongated shape along the axis of the loop. If there is a source in the plasma that emits a wave close to the plasma frequency in all directions, the emission when leaving the generation region, becomes acutely directed.

This leads to the fact that the registered emission, due to the curvature of the DPR layers, comes from very small areas near the tube axis and substantially larger areas (by 1-6 orders of magnitude) at a certain distance from the tube axis (of the order of the radius). These layers are already located almost parallel to the axis of the tube. The distance between the centers of the emitting regions for the harmonic numbers $s$ and $s + n$ in this model is (Karlický and Yasnov, 2019):

$$dh = -\frac{L_{bh} \ln\left[\frac{-1+(n+s)^2}{-1+s^2}\right]}{-2 + L_{br}^2} = \frac{R\, L_{nh} \ln\left[\frac{-1+(n+s)^2}{-1+s^2}\right]}{2 - R}, \quad (5)$$

where $R = L_{bh}/L_{nh}$.

## 2.1. Method for Calculating the Harmonic number of the Selected Stripe and the Ratio of the Field and Density Scales along the Axis of the Radiating tube.

When diagnosing physical conditions in generation regions of bursts with ZP, it is important to determine the value of the harmonic number $s$ corresponding to some stripe in the time-frequency spectrum. In (Karlický and Yasnov, 2015), a technique was developed for determining the harmonic number $s$ of the selected stripe in the spectrum of a microwave burst with ZP. In what follows we offer a working procedure for calculating the harmonic number ($s1$) for given a zebra stripe. It is preferable to carry out the frequency approximation of the zebra stripes by a polynomial of the second degree – $f(n)$, where $n$ is the serial number of the zebra stripe with a frequency $f$. This approximation will immediately make it possible to determine whether the harmonic number

Alternative Models...

decreases or increases with increasing frequency. With a positive coefficient of a quadratic term, the harmonic number decreases with increasing frequency and vice versa. This trend can be determined in advance and by the method presented in (Karlický and Yasnov, 2018b). Further, it is necessary to minimize the relation 11 from (Karlický and Yasnov, 2018b):

$$\left| (f(1+|d|)/f(1))^2 - \frac{(-1+s1^2)(\frac{\sqrt{-1+(d+s1)^2}}{\sqrt{-1+s1^2}})^{-L_{nb}}(d+s1)^2}{s1^2(-1+(d+s1)^2)} \right| + \left| (f(N)/f(N-|d|))^2 - \frac{(-1+(s1+ds)^2)(\frac{\sqrt{-1+(d+s1+ds)^2}}{\sqrt{-1+(s1+ds)^2}})^{-L_{nb}}(d+s1+ds)^2}{(s1+ds)^2(-1+(d+s1+ds)^2)} \right|, \quad (6)$$

where $ds=s2-s1$, and $s1$ is the harmonic number of the selected stripe, $s1 + d$ and $s2 + d$ are the harmonic numbers of the additionally selected stripes. If $s2 + d$ corresponds to the stripe with the highest frequency (with number $N$), we have $ds=\pm(N-|d|-1)$. Here, the sign is opposite to the coefficient sign for the quadratic term of the approximation mentioned above. If $s1$ corresponds to the lowest frequency, the negative values of $ds$ and $d$ mean a decrease in the harmonic number with frequency, and positive values mean an increase in the harmonic number with frequency. These signs are determined with high reliability in the process of such calculations, since the residual for different signs differs by several orders of magnitude.

The relation 6 can be simplified by replacing the parameter used there by the scale ratio $R = L_{bh}/L_{nh} = \frac{2L_{nb}}{2+L_{nb}}$. Then this expression will have the following form:

$$\left| (f(1+|d|)/f(1))^2 - \frac{(\frac{-1+(d+s1)^2}{-1+s1^2})^{\frac{2}{R-2}}(d+s1)^2}{s1^2} \right| + \left| (f(N)/f(N-|d|))^2 - \frac{(\frac{-1+(d+s1+ds)^2}{-1+(ds+s1)^2})^{\frac{2}{R-2}}(d+s1+ds)^2}{(s1+ds)^2} \right|. \quad (7)$$

We note that the values of $s1$ and $R$ are determined uniquely whether the emission occurs at the upper hybrid frequency or at its second harmonic. It was shown in (Karlický and Yasnov, 2018b) that the centers of the emitting regions in the selected time-frequency stripes should be at the same distance $r$ from the tube axis.

**2.2. The Burst Analysis with ZP Observed on 21 June 2011**

We have carried out the corresponding calculations for the burst with ZP that was analyzed in (Kaneda *et al.*, 2018). This work presents data on the detection of short-period propagating fast sausage mode waves in a metric radio spectral fine structure observed on 21 June 2011 with the Assembly of Metric-band Aperture Telescope and Real-time Analysis System. Analysis of zebra patterns in a type-IV burst revealed a quasi-periodic modulation in the frequency separation between the adjacent stripes of the ZPs (Δ*f*). Based on the double plasma resonance model, the most accepted generation model of ZPs, the observed quasi-periodic modulation of the ZP can be interpreted in terms of fast sausage mode waves propagating upward at phase speeds of 3000–8000 km s$^{-1}$. In this case, the application of the DPR mechanism was based on the presence of strong polarization in the ordinary mode and on Δ*f* increase with emission frequency. But any radioemission near the plasma frequency is likely to be polarized in the ordinary mode, and a decrease in Δ*f* with an increase in emission frequency is also possible when using the mechanism with the DPR (Karlický and Yasnov, 2018b). The authors' calculations were based on the Baumbach – Allen formula for plasma density (Allen 1947) and the Dulk and Mclean formula for magnetic field (Dulk and McLean 1978), and it was assumed that the characteristic spatial scale of the plasma density is greater than that of the magnetic field ($|L_{nh}| > |L_{bh}|$). It was also assumed that the magnetic field strength in the ZP generation region is 10 G. Below we are showing that if we base our considerations on assumptions similar to the ones in (Kaneda *et al.*, 2018), that is if we assume that the double plasma resonance model is valid and that radio emission occurs at the first harmonic of the upper hybrid frequency, then the field in the burst generation region is expected not to exceed 1-1.5 G. Therefore, the velocities of MHD waves should be much lower and a similar interpretation of a quasi-periodic modulation in the frequency separation between the adjacent stripes is untenable. The field in the source might be supposed to be close to 10 G, but then the ZP generation mechanism cannot be associated with the DPR.

Using the methodology described above, we determine the parameters of the generation region of this ZP, assuming, as in (Kaneda et al. 2018), that the ZP emission is associated with the DPR and occurs at the first harmonic of the upper hybrid frequency. Figure 1 shows the spectrum of this burst, on which the stripe with the harmonic number *s* = 55 is marked, as will be determined below.

Alternative Models. . .

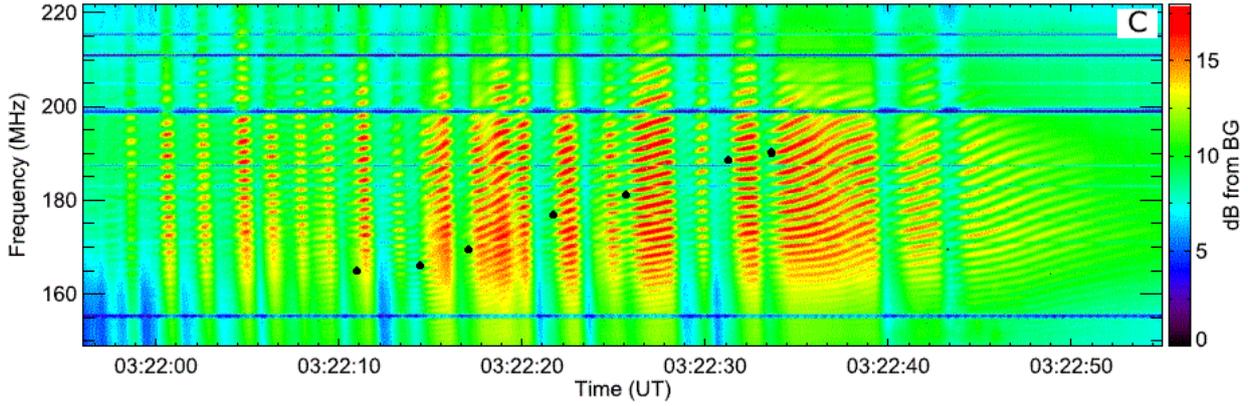

Figure 1 The burst spectrum on 21: 06: 2011 from (Kaneda et al. 2018), in which the stripe with the harmonic number $s\approx 55$ is marked with a black spot.

Table 1 shows the parameters of the burst generation region for 2011 June 21 at different points of time under the assumption that the ZP emission is associated with the DPR. In it $s1$ is the harmonic number for the stripe with the lowest frequency; $s_{fix}$ is the harmonic number for some fixed stripe. The values of the magnetic field strength $B$ is given in the generation region of the stripe at $f$=183 MHz. The ratio of the scales of the magnetic field and the density along and across the axis of the emitting region is $R=L_{bh}/L_{nh} = (L_{br}/L_{nr})^2$, and the relative distance $dh/L_{nr}$ between the levels of the DPR for $s$ and $s$ +1 at $f$=183 MHz. are also given.

Table 1. Parameters of the burst generation region at different moments of time for 2011 June 21. f are the frequencies of the extreme stripes ZP, $s1$ is the harmonic number for the stripe with the lowest frequency, $s_{fix}$ is the harmonic number for some fixed stripe. Magnetic field strengths $B$ for the stripe near the frequency of 183 MHz, the ratio of the scales of the magnetic field and the density along and across the axis of the emitting region $R =L_{bh}/L_{nh} = (L_{br}/L_{nr})^2$, the relative distance $dh/L_{nr}$ between adjacent levels of the DPR emitting at a frequency near 183 MHz are presented.

| Time | $s1$ | $R= L_{bh}/L_{nh}$ | $s_{fix}$ | $s$ at $f$=183 MHz | $B$, G for $f$=183 MHz | $dh/L_{nr}$ for $s$ и $s$ +1 at $f$=183 MHz |
|---|---|---|---|---|---|---|
| 3:22:11.39 | 50.47 | 0.632 | 64.47 | 44 | 1.485 | 2.079 |
| 3:22:18.97 | 59.63 | 0.73 | 66.63 | 50 | 1.307 | 2.280 |
| 3:22:22.57 | 61.87 | 0.729 | 64.87 | 53 | 1.233 | 2.145 |
| 3:22:27.42 | 70.44 | 0.792 | 70.44 | 55 | 1.188 | 2.33 |
| 3:22:32.35 | 74.49 | 0.835 | 74.49 | 58 | 1.127 | 2.452 |
| 3:22:35.37 | 69.19 | 0.843 | 68.19 | 59 | 1.108 | 2.451 |
| 3:22:38.23 | 71.98 | 0.882 | 69.98 | 61 | 1.071 | 2.566 |
| 3:22:41.91 | 69.99 | 0.885 | 66.99 | 63 | 1.037 | 2.501 |

Let us analyze the error in determining $s1$. First, we consider a possible methodological error related to the method for determining $s1$ from the minimum of the residual of the relation (7). Figure 2 shows the residual values $D$ depending on the variable fixed value $s1$. It can be seen that the discrepancy for $s1 = 70.44$ is more than an order of magnitude smaller than for the nearest values $s1 = 70$ and 71. Therefore, it can be argued that this method for determining $s1$ gives its values with very high accuracy.

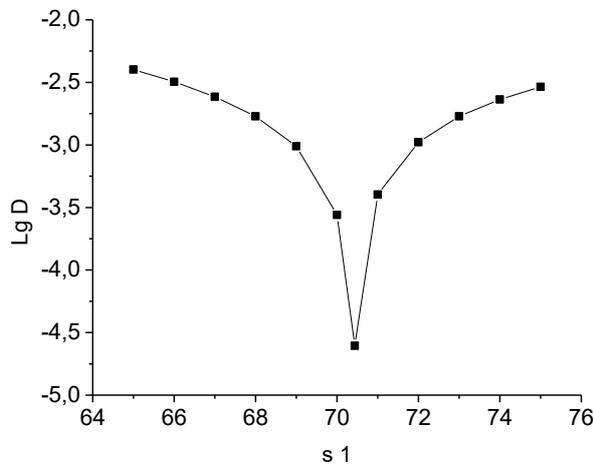

Figure 2 The residual value $D$, depending on a fixed value $s1$.

Let us determine the possible error in determining $s1$ associated with the error in determining the frequencies of the emission maxima of the emission stripes of the ZP. To do this, according to the formula for the frequency of the stripe 11 from (Karlický and Yasnov, 2019):

$$f_{UH}^s = 2.8\, s \left(\frac{s^2-1}{s1^2-1}\right)^{\frac{1}{R-2}} B(s1) \quad (8)$$

for different $s$, we calculate the frequencies of the corresponding stripes and add some value of the random error. Our error in determining of the values of the emission maxima of the stripes is about 0.3%. In accordance with previous calculations, we set $R = 0.75$, $s1 = 65$, $B(s1) = 1$ G. Then, with a zero error, we get $s1=64.95$ while with an error of 0.3%, we get each time new values within the range $s1=57/77$. The spread of the measured values is somewhat smaller but close to this spread. So we can conclude that the spread of the values we obtained for $s1$ is completely determined by the error of our determining the stripes frequencies in this spectrum.

One can assume confidently that at such low frequencies a burst occurs in the coronal region. Therefore, when assigning $L_{nh}=50\, T[K]/L_{nr}$ (Priest, 2014), we will assume that the temperature in the region of burst generation is close to $T = 2\times10^6$ K. Figure 3 shows the dependence of $R$

Alternative Models...

$=L_{bh}/L_{nh}$ and $L_{br}/L_{nr}$ on time. Figure 4 shows the field, as well as the distance between the stripes relative to the density scale along the radius, $dh/L_{nr}$ at a frequency of 183 MHz as functions of time. Figure 5 shows the change in time of the distance between the stripes with $s = 55$ and $56$ and the neighboring stripes near the frequency ≈183 MHz relative to the density scale along the radius. It can be seen that the relative distance $dh/L_{nr}$ between adjacent stripes near 183 MHz varies less than the distance between the stripes with $s = 55$ and $56$.

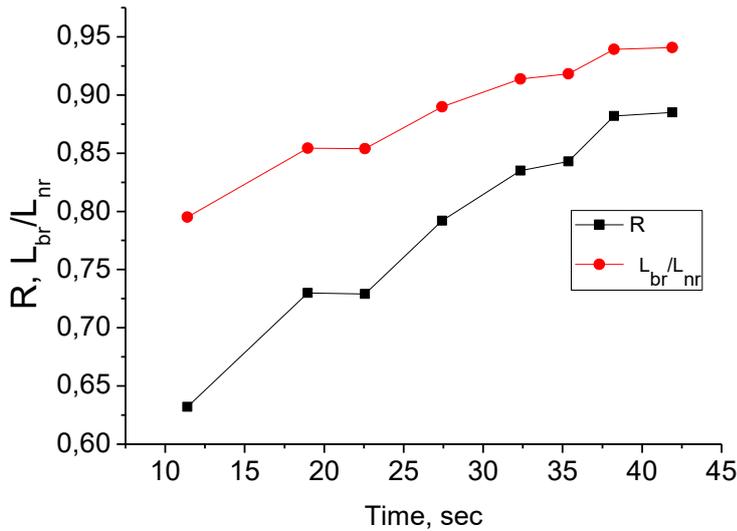

Figure 3 The dependence of $R = L_{bh}/L_{nh}$ and $L_{br}/L_{nr}$ on time. Hereafter, the time is counted from 3:22:00 UT.

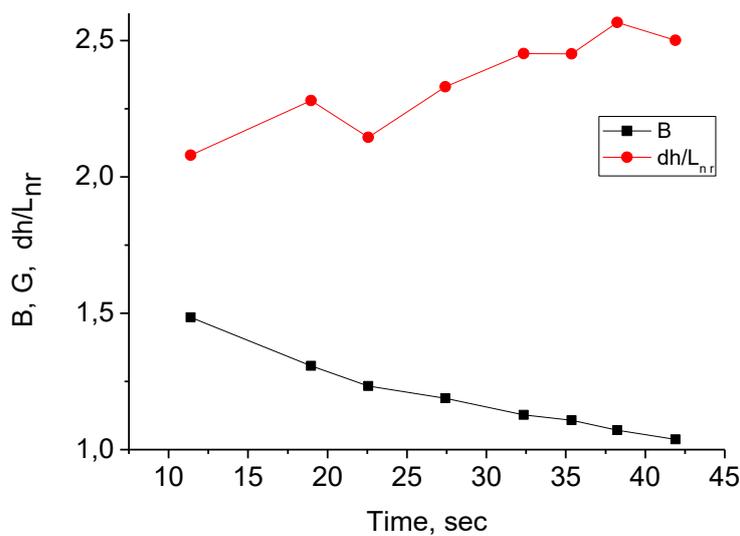

Figure 4 Field $B$ and $dh/L_{nr}$ - the distance between the stripes at a frequency of 183 MHz relative to the density scale along the radius.
9

A feature of this burst is that the frequency interval where the ZP stripes are generated does not change over time. Obviously, this is because the density in the generation region of the given ZP does not depend on time.

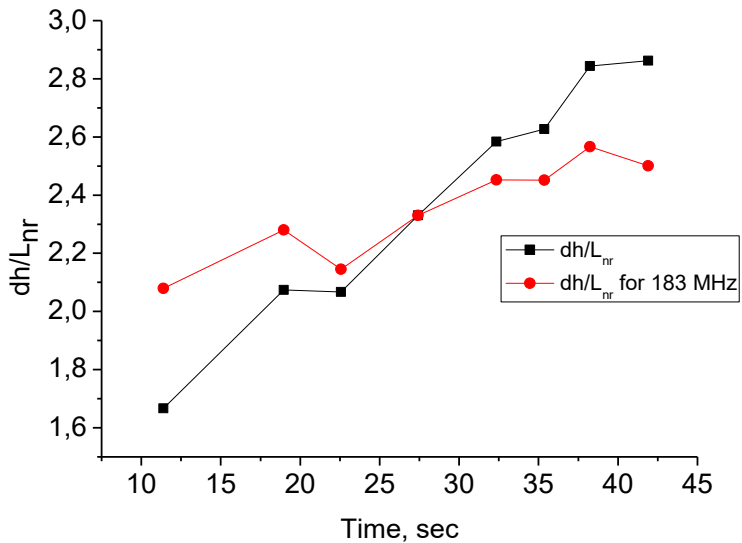

Figure 5 The change over time of the distance relative to the density scale along the radius between the stripes with $s$ = 55 and 56 and the stripes near the frequency ≈183 MHz.

## 3. Discussion of the Results of Calculations Performed in the Framework of the Model with DPR

Corresponding calculations are presented for the burst with ZP that was analyzed in (Kaneda et al. 2018), within the framework of the generation mechanism based on the double plasma resonance model. It is shown that the parameter $dh/L_{nr}$, that describes the distance between the regions of stripe generation relative to the density scale along the radius $L_{nr}$, increases with time. The generation region of this ZP is limited in frequency. In this case, the density in the generation region is practically unchanged, and the magnetic field strength changes from 1.485 to 1.037 G during a burst of 35 seconds. The distance between the stripes with $s$ = 55 and 56 changes 1.7 times, while the distance between the nearest stripes near the frequency ≈183 MHz changes 1.5 times.

In the meter range, the emission is more likely to occur at the second harmonic (Karlický, Yasnov, 2018a). In the second harmonic, the emission can exit in substantially larger directions and

Alternative Models. . .

the region of possible emission of an individual spectral stripe increases many times (Karlický, Yasnov, 2019). So, the sizes of the radiating regions near the straight tube along its axis at the second harmonic for $s = 55$ are 57 times larger than the sizes of the radiating regions at the first harmonic, regardless of the region position along $r$. In this case, the values of $s$ and $R$ determined by the above procedure stay unchanged. Then the magnetic field is reduced by 2 times. And at the same time there is no explanation for the predominant emission in the ordinary mode. In addition, since the size of the emission region increases many times, significant difficulties arise in the generation of isolated ZP stripes, because in this case the condition 4 in a large emission region can be violated.

If we base our considerations on the mechanism with DPR, another question arises. How can it be explained that the field strength in the ZP generation region changes with time from 1.485 to 1.037 G, while the density is practically unchanged? Taking into account the correlation (or equality) of magnetic and gas pressures, this is not to be expected. In addition, in the solar atmosphere, the field and the density tend to decrease in a correlated way with altitude. Besides, such changes result in fulfilling strictly the relation 3, which is also difficult to imagine.

We also note that, for the values given above, that is, for the plasma frequency $f_P = 183$ MHz, $B \approx$ 1.2 G and $T = 2$ MK, the plasma betta is equal $\beta = 2$. And for the emission at the second harmonic $\beta = 8$. Such values are quite unlikely, since it is known that the plasma betta in the corona should be significantly less than 1. In addition, it was noted in (Kolotkov, Nakariakov, and Kontar,. 2018) that for type III bursts, the plasma frequency of 34 MHz takes place where the magnetic field is $B \approx 1.1 \pm 0.2$ G. That is, it is unlikely that the field obtained above (1-1.5 G), and even more so the field two times less would correspond to the plasma frequency of 183 MHz.

We also note other possible difficulties in the mechanism of generation of bursts with ZP based on the DPR. For example, in (Karlický, Yasnov, 2015) it was shown that in an atmosphere with a barometric dependence of the electron density on the height, the radio emission at frequencies corresponding to the upper hybrid frequency experiences strong absorption, and such a burst cannot be detected. In (Yasnov, Benáček, Karlický, 2019) it was shown that the increment of the upper hybrid waves has isolated maxima only in a rather narrow interval of the pitch-angle boundary (about 50 degrees) in the distribution of high energy electrons. For small angles the growth rate is broad and flat and for high pitch angles even absorption occurs. Moreover, in all cases a sharp low-energy boundary is required in the distribution of high-energy electrons. The origin of such a sharp boundary is unknown. Also in the paper (Benáček, Karlicky, Yasnov, 2017) for the DGH function of high energy electrons distribution (Dory, Guest, and Harris, 1965) and by taking into account relativistic corrections, it was shown that a clear increment maximum is obtained only for the electron velocities of 0.1 $c$ with a narrow dispersion. For a velocity of 0.2 $c$, the increment decreases sharply and the maxima disappear even for relatively small values of cyclotron harmonics $s$.

The above reasoning shows that the generation mechanism of this zebra structure is unlikely to be based on the DPR In this regard, we consider another possible mechanism for the formation of zebra stripes, based on the interaction of plasma waves with whistlers.

## 4. Model with Whistlers

The model for the ZP formation that takes into account the interaction of plasma waves with whistlers was developed almost simultaneously with the DPR model, starting with the work of Chernov, 1976. The main details of the model with whistlers are considered in a number of reviews and monographs (Chernov, 2006; 2011).

It should be noted that all works discussing the mechanism of DPR (and its improvement), do not take into account an important property of fast electron distribution functions (with a loss-cone, ring distribution or with a power-law spectrum), that is, the fact that whistlers waves are excited simultaneously with the plasma waves, so the basic dynamics of the distribution function and the radio emission is determined by the interaction of fast particles with the whistlers (Chernov, 1996). As a result, the magnetic trap (the source of the zebra structure) is filled with periodic whistler gain zones separated by their absorption zones. The length of the gain zones along the trap is $\approx 10^8$ cm, the zones being separated by damping zones of approximately the same size. Amplification zones (whistler wave packets) move with a group velocity, which results in the regular stripes of the zebra structure being excited. The generation of periodic whistler wave packets by anisotropic particle beams was analyzed theoretically and was confirmed by numerous laboratory experiments (Kuipers, 1975, Chernov, 1976).

The magnetic field in the model with the whistlers is determined by the frequency separation between the centers of stripes in the emission and absorption spectra (equal to the frequency of whistlers), which is approximately 2 times less than the frequency separation between the stripes in the emission $\Delta f$. The frequency of whistlers should be taken not at the maximum of the group velocity ($\approx \frac{1}{4} f_B$), but at the maximum of the excitation increment of whistlers, which is located on the level $\approx 0.1 f_B$ (for example, see Maltseva and Chernov, 1989). In 21.06.2011 event in Figure 2 the frequency separation between the stripes in the emission at the frequency 180 MHz at the moment 02:22:35 UT is of 2.5 MHz. Where from we obtain $f_B$ =12.5 MHz, and B = 4.46 G. For this field strength we obtain the plasma betta $\beta$ = 0.14, which corresponds completely to the conditions

Alternative Models. . .

in the source of the magnetic trap type. This value of the magnetic field strength is sufficiently close to the utilized value in the paper (Kaneda *et al*., 2018).

The second important issue in the comparative analysis of the mechanisms is the estimation of the possible number of simultaneous zebra stripes. The issue should be considered in two aspects: firstly, what models of the magnetic field and concentration in the corona would provide the maximum number of stripes, and secondly, what number of stripes the generation mechanism would be able to give.

As mentioned above, the mechanism DPR for the ring-function of distribution is very limited in the number of increment maxima and in the particle velocities. As Kuznetsov and Tsap (2007) showed, a large number of stripes are realized only with a power-law spectrum of hot electrons with a spectral index of about 8. According to RHESI data, such steep spectra are very rare even in large phenomena. Algebraically, such high harmonics as $s$ = 55-57 are easily obtained from the DPR formulas. However, the increment calculations performed there are limited to a rare steep spectrum of fast particles. In addition, in the known models of the magnetic field and concentration, a large number of DPR levels are not realized (see, for example, Figure 21 in Chernov (2010)).

The whistler mechanism operates in a wide range of the scales heights relationships of the magnetic field and density. When a magnetic trap with dimensions > $10^9$ cm is filled up with wave whistler wave packets with scales ≤ $10^8$ cm, several tens of stripes can be simultaneously excited. Thus, in the whistler model, there is no need to find the number of some harmonic for a particular zebra stripe, as it is necessary to find in the DPR model ($s$ = 55 in Figure 1).

In the DPR mechanism, all non-stationary changes in the zebra- stripes (wave-like behavior of the frequency drift) are explained by a change in the magnetic field, or by a passage of a fast magnetosonic wave. More complex spectra with a stripes splitting and superfine structure are not considered at all or are explained by hypothetical plasma turbulence in the source.

Whistlers are excited by anisotropic particle beams. When the pulse injection of beams in the bases of trap occurs, the whistlers are excited on the normal Doppler Effect and with the predominance of transverse anisotropy, they move upward and disperse in space away from the fast particles. There are no quasi-linear effects of their interaction, and mechanism on the whistlers excites stripes with a constant negative frequency drift of the fiber type (fiber bursts).

The prolonged particle injection causes the runaway electrons to modify the distribution function with the predominance of particles with the longitudinal velocities. The whistler instability switches to the anomalous resonance. In this case, particles and waves propagate in almost the same direction, and the quasilinear interaction is activated. In the further dynamics of the distribution function, the whistler instability can switch from the anomalous to the normal Doppler effect. Since the whistler group velocities in the normal and anomalous effect have opposite directions, the frequency drift of the resulting zebra stripes should change to the opposite one. For a smooth dynamics of the distribution function, a wave-like change in the frequency drift is to be observed. For a sharp change in the beam anisotropy (for example, with a pulsed additional injection of particles), a sharp change in the frequency drift is to be observed while the zebra stripes would have a sawtooth, character, which is often observed (for more details see Chernov (1996)).

Whistlers that propagate propagating down to the photosphere from the excitation region decay rapidly due to cyclotron damping when the whistler frequency approaches to half of the cyclotron frequency. Whistlers can propagate upward into the corona without attenuation along the entire magnetic trap. Therefore, the frequency band with the zebra structure is actually determined by the size of the magnetic trap. In the DPR model, a problem arises how to explain a large number of stripes (in the event under consideration, up to 27 bands in Figure 2 at 03:22:28 UT). According to the algebraically selected harmonic number $s = 55$ for the selected stripe, the number of stripes should be associated with harmonic numbers from 43 to 70. But, for example, for the DGH distribution (Dory, Guest and Harris, 1965) the maximum growth rates of excited harmonics in the DPR model are obtained only for the first few harmonics (Benáček, Karlický and Yasnov, 2017).

The main feature of the mechanism on the whistlers is considered to be a synchronous sign change of the stripes frequency drift on the spectrum and directions of the spatial drift of their radio sources. Similar observations are sure to be realized by using the new Chinese radio-heliograph MUSER, which possesses sufficient spatial and time resolutions.

A splitting of a stripe into two ones with a small frequency separation can be observed at the moments when instabilities of the normal and anomalous effects switch on simultaneously, the effect being observed in (Chernov *et al*., 1998). In the event under consideration, the effect occurs as well. The most obvious case of splitting can be seen in Figure 2A in (Kaneda *et al*., 2015) in 03:22:40 UT near the frequencies 171 MHz.

This event shows no explicit superfine (millisecond) structure of stripes or it might not have been investigated. In the decimeter and microwave ranges such structure is always observed, which agrees with the model with the whistlers. The millisecond structure of strips is formed as a result of the interaction of whistlers with the ion-acoustic waves in the pulsating regime. The necessary condition for this to happen (temperature anisotropy) does not always realize at the heights of a meter range.

The DPR level on the harmonic $s$=55 is unlikely to remain constant for 30 seconds, in spite of pulsation in the absorption, i.e., additional particle injections. In the model with the whistlers the

Alternative Models. . .

source is also distributed along the height, but all components of stripes are explained by natural processes of the whistlers' propagation with the group velocity as well as by their interaction with the fast particles, the explanation not involving any turbulence. Thus, the model with the whistlers is more suitable for explaining the parameters of the phenomenon 21 June 2011 than the model on DPR.

## 5. Conclusions

We have considered two alternative models for the formation of a zebra structure in the event of 21 June 2011. One of them is based on the DPR (as in Kaneda et al. (2018)), the other is based on the whistler model. In the case of the DPR, the ratio of the scale characteristics of the field and density along the axis of the power tube $L_{bh}/L_{nh}$ and across its axis $L_{br}/L_{nr}$ as well as their dependence on time are analyzed. It has been shown, for example, that the relative distance between adjacent stripes $dh/L_{nr}$ near 183 MHz varies less than the distance between the stripes with $s = 55$ and 56. In (Kaneda *et al.,* 2018), it was assumed that the field strength in the source is about 10 G. But in accordance with our calculations, the magnetic field strength turned out to be almost an order of magnitude lower and, accordingly, the speed of fast sausage mode waves that may be present should be almost an order of magnitude lower as well. Therefore, the cause of the effect detected in (Kaneda *et al.*, 2018) should be sought either by not basing the reasoning on MHD waves, or by changing the mechanism for generating stripes of this ZP.

At the meter range, emission at the second harmonic is more probable, since emission can escape in substantially more different directions and the size of the region of a possible emission of an individual spectral stripe increases many times. Even so, a problem arises related to the possible merging of the stripes due to the difficulty of fulfilling the condition (4). The magnetic field in this case is reduced 2 times. However, there is no explanation for the predominant emission in the ordinary mode.  There are a few more problems. There is difficulty in explaining the fact that the field strength in the ZP generation region changes with time almost one and a half times (from 1.485 to 1.037 G), while the density is practically unchanged? Another difficulty for this mechanism lies in the fact that emission at the first harmonic gives the plasma beta $\beta \approx 2$, while emission at the second harmonic gives $\beta \approx 4$. Such values are unlikely, since the plasma beta in the corona is known to be noticeably less than 1.

The model with whistlers at a qualitative level explains all the main observational characteristics of this zebra. Its main advantage is that for its implementation in this event, a significantly larger magnetic field (4.5 G) is needed. For such a field value, the value of the plasma beta is $\beta = 0.14$, which fully corresponds to the coronal conditions and conditions in the source of the type of magnetic trap. The fluctuations $\Delta f$ are simply explained by fluctuations in the direction of the group velocity of the whistlers, as evidenced by the wave-like oscillations of the frequency drift of the stripes. In addition, the frequency of the whistlers changes depending on the propagation angle (without involving hypothetical MHD waves).

**Acknowledgments**

L.V. Yasnov acknowledges support from the Russian Foundation for Basic Research, Grants 18-29-21016-mk, 19-52-26016-Czech-a and partly from Grant 18-02-00045, from Program RAN №28, Project 1D and State Task № AAAA-A17-117011810013-4. G.P. Chernov acknowledges support from the Russian Foundation for Basic Research, Grants 17-02-00308 and from Program № 12 of the Presidium RAN.

**Disclosure of Potential Conflicts of Interest** No potential conflict of interest was reported by the authors.

Alternative Models. . .